\documentclass[aps,pra,twocolumn,superscriptaddress,longbibliography]{revtex4-2}

\usepackage{amsmath,amssymb,amsfonts}
\usepackage{graphicx}
\usepackage{bm}
\usepackage{braket}
\usepackage{xcolor}
\usepackage[colorlinks=true,
            linkcolor=blue,
            citecolor=blue,
            urlcolor=blue]{hyperref}
\usepackage{dcolumn}
\usepackage{array}
\newcolumntype{d}[1]{D{.}{.}{#1}}

\newcommand{\lv}{\bm{\lambda}}
\newcommand{\He}{\hat{H}_{\rm e}}
\newcommand{\HPF}{\hat{H}_{\rm PF}}
\newcommand{\HCS}{\hat{H}_{\rm CS}}

\newcommand{\wcav}{\omega_{\rm c}}
\newcommand{\lam}{\lambda}

\begin{document}

\title{Cavity-modified static polarizabilities of atoms and small molecules within QED full configuration interaction}

\author{T. Zalialiutdinov}
\affiliation{Department of Physics, St. Petersburg State University,
Petrodvorets, Oulianovskaya 1, 198504 St. Petersburg, Russia}
\affiliation{Petersburg Nuclear Physics Institute named by
B.~P.~Konstantinov of National Research Centre ``Kurchatov
Institute'', Gatchina 188300, Russia}

\author{D. Solovyev}
\affiliation{Department of Physics, St. Petersburg State University,
Petrodvorets, Oulianovskaya 1, 198504 St. Petersburg, Russia}
\affiliation{Petersburg Nuclear Physics Institute named by
B.~P.~Konstantinov of National Research Centre ``Kurchatov
Institute'', Gatchina 188300, Russia}

\date{\today}

\begin{abstract}
Static dipole polarizabilities of H$_2$, LiH, and helium confined in
a single-mode optical cavity are computed using
quantum-electrodynamical full configuration interaction (QED-FCI).
Response properties are obtained as finite-field derivatives of the
coupled electron-photon energy. The cavity-induced modification is
found to arise from the competition of two contributions of opposite
sign, the dipole self-energy and photon exchange, whose relative
strength depends on both the cavity frequency and the number of
photons involved in the exchange. The residual change suppresses the
response of the molecular ground electronic states. For the
metastable states of helium atom this balance changes qualitatively, and
the differential response of the $2\,^3S_1 \to 2\,^1S_0$ clock
transition, which governs its sensitivity to static electric fields
and to thermal radiation, is strongly enhanced instead. Cavity
confinement can therefore either reduce or increase the dc Stark and
blackbody-radiation shifts of a frequency standard, depending on the
electronic structure of the states involved. The reported data serve
as a reference for cavity-modified response and underscore the need
for an exact treatment of light-matter correlation in cavity-QED
predictions.
\end{abstract}
\maketitle

\section{Introduction}
\label{sec:intro}

The coupling of atoms and molecules to the quantized field of an
optical cavity has emerged as a practical means of modifying matter
without external
driving~\cite{Ruggenthaler2018,Flick2017pnas,Mandal2023}. When the
coupling strength approaches electronic energy scales, the electronic
and photonic degrees of freedom become
entangled~\cite{Haugland2020jctc,Riso2022}, and as a consequence
ground-state observables such as the geometry, the dipole moment, and
the electron density are altered already by vacuum
fluctuations~\cite{Flick2017pnas,Horak2025,Philbin2023}.
The static dipole polarizability is a natural target for this kind of
control. It governs the linear optical response, sets the sensitivity of
a system to stray electric fields, and enters long-range dispersion
coefficients~\cite{Philbin2023,DePrince2025}. In precision metrology it is
often this quantity that limits accuracy. The blackbody radiation shift
of a clock transition is proportional to the differential static
polarizability of the two clock states~\cite{Itano1982,Safronova2012},
and the magic wavelength of an optical lattice is defined by the
vanishing of its dynamic counterpart~\cite{Zanon2006}. Blackbody
radiation also drives the photodissociation of the trapped molecular ions
used in some clock proposals through the same response, so that a
cavity-induced change would affect both the frequency and the lifetime
of such a standard.

Existing studies of cavity-modified molecular response have been carried out 
mostly at the mean-field level or within density functional 
theory~\cite{Horak2025,Philbin2023}, while correlated treatments predominantly 
rely on quantum electrodynamics coupled cluster (QED-CC) 
theory~\cite{Haugland2020jctc,Haugland2021jcp,DePrince2021}. However, the accuracy 
of truncated QED-CC methods for response properties remains to be fully 
characterized. Near the equilibrium geometry, its accuracy is primarily limited 
by the truncation of the photon sector. Near bond dissociation, an additional 
and more severe issue arises because the single-reference description on which 
the method is built may break down. This effect can be further compounded by 
the photon basis truncation. Even for a well-behaved reference, such a limitation 
was recently demonstrated for a two-electron system where the electronic part 
of the ansatz is otherwise exact. In that case, QED-CCSD with two photon 
excitations was shown to underestimate electron-photon 
correlation~\cite{Perspective2025}. Reference data free of both sources 
of error are therefore highly desirable.

In this work, the QED full configuration interaction (QED-FCI)
approach is employed, in which the electron-boson Hamiltonian is
diagonalized directly in the product space of electronic Slater
determinants and photonic Fock states. Within a given one-particle
basis and photon-number truncation, such a diagonalization is
numerically exact, so that electronic and electron-photon correlation
are treated on the same footing and no particular ansatz for the form
of the light-matter entanglement has to be assumed. Three systems are
considered in the following. As a first, nonpolar benchmark, the
dipole polarizability of H$_2$ is computed, while for the polar
molecule LiH the permanent dipole moment is evaluated alongside the
polarizability. The study is then extended to helium atom, where not only
the ground state but also the two lowest metastable excited states are
investigated. Throughout, all properties are obtained by the
finite-field method~\cite{Cohen1965}, in which the total energy is
recomputed in the presence of a weak static electric field and the
response properties are extracted by numerical differentiation with
respect to the field strength.

In this work we analyze how the polarizability of an atom or molecule
is modified when it is embedded in a cavity, and identify two
competing contributions of opposite sign underlying this effect. The dipole
self-energy is independent of the cavity frequency and invariably
suppresses the polarizability along the mode axis. Photon exchange, in
contrast, acts in the opposite direction, with a magnitude governed by
the cavity frequency. For the molecular ground states these two
contributions largely cancel. The cancellation is incomplete, however,
and a residual suppression of the order of ten percent nonetheless
remains. For the metastable states of helium the situation is
qualitatively different, since there the cancellation does not occur
and the polarizability instead increases substantially. Both regimes
are quantified through their consequences for the blackbody radiation
shift, which thus emerges as a sensitive probe of the balance between
the two competing contributions.

The paper is organized as follows. In Section~\ref{sec:theory}, the
Pauli--Fierz Hamiltonian is introduced together with the
coherent-state transformation used for polar molecules, and the
finite-field procedure employed to extract the dipole moments and
polarizabilities is described. Section~\ref{sec:comp}
summarizes the computational details, including the choice of basis
sets, the truncation of the photon space, and the validation of the
implementation. The results are presented in Section~\ref{sec:results},
where the cavity-induced modification of the polarizability is
analyzed for H$_2$ and LiH and then extended to the ground and
metastable states of helium. Conclusions are drawn in
Section~\ref{sec:concl}. Atomic units, $\hbar=e=m_{e}=1$, are used throughout the
paper.
% =======================================================
\section{Theoretical background}
\label{sec:theory}
\subsection{Electron-photon Hamiltonian and the QED-FCI approach}
\label{pauli-fierz}
We consider a system coupled to one cavity mode of frequency $\wcav$ and
polarization $\hat{\varepsilon}$. Within the Born-Oppenheimer and dipole
approximations the electron-photon system is described by the
Pauli--Fierz Hamiltonian in the length
gauge~\cite{Ruggenthaler2014pra,Zalialiutdinov2025dboc},
\begin{eqnarray}
\label{eq:HPF}
\HPF = \He + \wcav\hat{b}^\dagger\hat{b}
       - \sqrt{\frac{\wcav}{2}}\,(\lv\cdot\hat{\mathbf{d}})(\hat{b}^\dagger+\hat{b})
       + \tfrac{1}{2}(\lv\cdot\hat{\mathbf{d}})^2 .
\end{eqnarray}
where $\hat{b}^\dagger$ and $\hat{b}$ denote the photon creation and annihilation 
operators. The standard electronic Hamiltonian $\He$ for a system of $N_e$ electrons 
and $N_n$ nuclei reads
\begin{eqnarray}
\label{eq:He}
\He = \sum_{i=1}^{N_e}\!\left(-\tfrac{1}{2}\nabla_i^2
     - \sum_{A=1}^{N_n}\frac{Z_A}{|\mathbf{r}_i - \mathbf{R}_A|}\right)
     + \sum_{i<j}\frac{1}{|\mathbf{r}_i-\mathbf{r}_j|}
\\\nonumber
     + \sum_{A<B}\frac{Z_A Z_B}{|\mathbf{R}_A-\mathbf{R}_B|},
\end{eqnarray}
and $\hat{\mathbf{d}} = -\sum_i \mathbf{r}_i + \sum_A Z_A \mathbf{R}_A$ is the total dipole operator 
entering Eq.~\eqref{eq:HPF}. 

The interaction of the atomic system with the cavity is given by the last 
three terms in Eq.~(\ref{eq:HPF}). 
The coupling vector is given by
\begin{eqnarray}
\label{eq:lambda}
\lv = \sqrt{\frac{1}{\varepsilon_0 V}}\,\hat{\bm{\varepsilon}},
\end{eqnarray}
where $\hat{\bm{\varepsilon}}$ is the unit polarization vector of the cavity mode, $V$ is the effective mode volume, and $\varepsilon_0$ denotes the vacuum permittivity. In atomic units, this yields $|\lv|^2 = 4\pi/V$, so that $\lam = 0.01$~a.u.\ and $\lam = 0.1$~a.u.\ correspond to mode volumes of $V \approx 19$~nm$^3$ and $V \approx 0.19$~nm$^3$, respectively.

The last two terms of Eq.~\eqref{eq:HPF} play distinct physical roles. The
bilinear term exchanges energy between matter and field and carries the
frequency-dependent factor $\sqrt{\wcav/2}$. The dipole self-energy term
originates from the diamagnetic $A^2$ contribution after the
Power--Zienau--Woolley transformation~\cite{Ruggenthaler2014pra}. This term is
frequency independent and is required for spectral positivity at strong
coupling. The qualitative difference between the two terms is the origin of
the two competing mechanisms discussed below.

The presence of the total dipole moment $\hat{\mathbf{d}}$ in the Hamiltonian
(\ref{eq:HPF}) makes the dipole self-energy term, and hence the energy obtained
in a finite one-particle basis, dependent on the choice of the origin.
Moreover, for a molecule with a permanent dipole moment, the dipole--field
coupling displaces the cavity mode from its vacuum state. It is therefore
convenient to work in the coherent-state representation, in which both effects
are removed by the unitary transformation
$\hat{U}(z)=\exp[z(\hat{b}^\dagger-\hat{b})]$ with
$z=\langle\lv\cdot\hat{\mathbf{d}}\rangle/\sqrt{2\wcav}$. Here and below, the
average is taken over the electronic state of the molecule, and $z$ is
determined self-consistently, as described below. In the transformed frame, the
self-energy term contains only the dipole fluctuation
$\hat{\mathbf{d}}-\langle\hat{\mathbf{d}}\rangle$, which is unchanged by a
constant shift of $\hat{\mathbf{d}}$, and the Pauli--Fierz Hamiltonian takes the
coherent-state form:
\begin{eqnarray}
\label{eq:HCS}
\HCS = \He + \wcav\hat{b}^\dagger\hat{b}
- \sqrt{\frac{\wcav}{2}}\,\bigl[\lv\cdot(\hat{\mathbf{d}} - \langle\hat{\mathbf{d}}\rangle)\bigr]
  (\hat{b}^\dagger+\hat{b}) 
\\\nonumber
+ \tfrac{1}{2}\bigl[\lv\cdot(\hat{\mathbf{d}} - \langle\hat{\mathbf{d}}\rangle)\bigr]^2
.
\end{eqnarray}

Since this transformation is unitary, the exact spectrum remains invariant with 
respect to $z$, which solely dictates the required size of the photon basis. 
In the untransformed frame, the ground state of a polar molecule is a coherent 
state with a mean photon number of $|z|^2$. For LiH at $\lam = 0.1$~a.u.\ and 
$\wcav = 0.0064$~a.u., this choice leads to a displacement of $|z| \simeq 2$ 
near the equilibrium bond length, which increases to $|z| \simeq 5$ at an elongated 
distance of $R = 4$~\AA. To ensure efficient photon basis truncation, we operate in 
the transformed frame and determine $z$ self-consistently from the correlated 
QED-FCI density, rather than relying on the QED-HF density used for the initial guess. 
Indeed, a Hartree--Fock estimate of $z$ becomes insufficient at large $R$. This stems 
from the well-known failure of restricted Hartree--Fock in the bond-stretching 
regime, where the wave function retains an unphysical ionic $\text{Li}^+\text{H}^-$ 
character, thereby overestimating the dipole moment.

Within QED full configuration interaction the wave function is expanded in the
product basis of Slater determinants and photon Fock states,
\begin{eqnarray}
\label{eq:fciansatz}
|\Psi\rangle = \sum_{I}\sum_{n=0}^{N_{\rm p}} C_{In}\,
|\Phi_I\rangle\otimes|n\rangle ,
\end{eqnarray}
and the Hamiltonian is diagonalized directly in this basis by the Davidson
algorithm~\cite{Knowles1984}. 
The coefficients $C_{In}$ can be organized into $N_{\rm p}+1$ blocks, one
for each photon number $n=0,1,\dots,N_{\rm p}$. Each block has the size and
structure of an ordinary electronic FCI vector. The electronic
Hamiltonian, the dipole self-energy, and the photon energy leave $n$
unchanged and therefore act within a single block. The bilinear term
changes $n$ by one and connects each block only to its two neighbors.

The calculation within a single block is thus identical to an ordinary
electronic FCI calculation, which requires only two arrays, a one-electron
matrix and a two-electron array $(pq|rs)$. Existing electronic structure
code can therefore be reused without modification, provided the dipole
self-energy and the coherent-state displacement $z$ are folded into these
arrays beforehand, using the dipole matrix
\begin{eqnarray}
\label{eq:dmat}
D_{pq} = -\lv\cdot\langle p|\mathbf{r}|q\rangle.
\end{eqnarray}
The one-electron matrix used throughout this work is
\begin{eqnarray}
\label{eq:hshift}
h_{pq} = h_{pq}^{(0)} + \tfrac{1}{2}\sum_r D_{pr}D_{rq}
- 2z\sqrt{\tfrac{\wcav}{2}}\,D_{pq},
\end{eqnarray}
where $h_{pq}^{(0)}$ is the bare one-electron integral, the second
term is the dipole self-energy, and the third term is the only
correction to $h_{pq}^{(0)}$ due to the displacement $z$. Here and below the geometry is centered on the center of nuclear charge,
so that the nuclear contribution to $\lv\cdot\hat{\mathbf d}$ vanishes. The two-electron
array, in contrast, is modified only by the self-energy,
\begin{eqnarray}
\label{eq:dse}
(pq|rs) \to (pq|rs) + D_{pq}D_{rs},
\end{eqnarray}
and is independent of $z$.

No further modification is required within a block. The photon energy
contributes a constant $n\wcav$ to block $n$, and the displacement $z$
contributes an additional constant, $\wcav z^2$, to the nuclear repulsion
energy. The bilinear term, which connects neighboring blocks, is constructed
from the matrix $-\sqrt{\wcav/2}\,D_{pq}$. The displacement $z$ adds
a further constant to this matrix as well, proportional to $z\wcav$.
As a result, the two-electron array is built once per geometry, and only
the one-electron matrix, the bilinear coupling, and a single constant are
updated during the self-consistency procedure described below. The only
approximations are the size of the one-particle basis and the truncation
$N_{\rm p}$ of the photon space, both of which are converged in the
present work.

\subsection{Finite-field evaluation of response properties}
\label{ssec:FF}

The polarizabilities and dipole moments are obtained numerically, by computing the total energy
in the presence of a small static electric field and inferring the response
from how the energy changes. This finite-field approach requires no
additional formalism beyond the ability to compute $E(\lambda)$ at a chosen
field strength, which is exactly what the machinery described above
already provides.

A static electric field $\mathbf{F}$ enters through the electronic contribution $+\mathbf{F}\cdot\sum_i\mathbf{r}_i$ and the nuclear dipole term $-\mathbf{F}\cdot\mathbf{d}_{\mathrm{nuc}}$. The total energy then depends on
$\mathbf{F}$ through the standard expansion
\begin{eqnarray}
\label{eq:expansion}
E(\mathbf{F}) = E_0 - \mathbf{d}\cdot\mathbf{F}
- \tfrac{1}{2}\sum_{ij}F_{i}\alpha_{ij}F_{j} + \ldots ,
\end{eqnarray}
where $E_0$ is the field-free energy, $\mathbf{d}$ is the permanent dipole 
moment, and $\alpha_{ij}$ denotes the polarizability tensor to be extracted.
Because $\alpha_{ij}$ is the matrix of coefficients of the quadratic term, it can be
obtained from $E(\mathbf{F})$ at a few small values of $\mathbf{F}$ using
finite differences. In the same way a second derivative of any smooth
function can be estimated from a handful of nearby function values. The displacement $z$ is determined once per geometry from the field-free
correlated density and kept fixed for all field points. Since the
transformation is unitary, the exact energies do not depend on this
choice, and the small field-induced change of the displacement is
accommodated by the photon basis.

In practice, a naive finite-difference formula converges slowly as the
field step $h$ is made smaller, because higher-order terms in
Eq.~\eqref{eq:expansion} contaminate the result at any finite $h$. This is
improved by evaluating the energy at two different step sizes, $h$ and
$2h$, and combining the two estimates of $\alpha_{ii}$ so that the leading
error cancels, a standard technique known as Richardson extrapolation.
Explicitly,
\begin{eqnarray}
\label{eq:richardson}
\alpha_{ii} = \tfrac{1}{3}\bigl[4\alpha_{ii}(h) - \alpha_{ii}(2h)\bigr]
+ \mathcal{O}(h^4),
\end{eqnarray}
which reduces the leading error from order $h^2$ to order $h^4$. In turn, the
difference $|\alpha_{ii}(h)-\alpha_{ii}(2h)|$ between the two estimates
provides a practical measure of how well converged the extrapolated value
is, and is reported throughout this work as an error estimate.

For a linear molecule with the cavity mode polarized along the molecular
$z$-axis, symmetry guarantees $\alpha_{xx}=\alpha_{yy}$, so only two independent
components, $\alpha_{zz}$ and $\alpha_{xx}$, need to be computed. Obtaining both, together with the extrapolation of
Eq.~\eqref{eq:richardson}, requires the energy at seven field points
in total. These are the field-free energy, four values along the
molecular axis ($\pm h$ and $\pm 2h$, needed for $\alpha_{zz}$), and
two values perpendicular to it ($\pm h$, needed for $\alpha_{xx}$). When the cavity mode is not aligned with the molecular
axis this symmetry is lost, all six independent tensor components
$\alpha_{xx}$, $\alpha_{yy}$, $\alpha_{zz}$, $\alpha_{xy}$, $\alpha_{xz}$,
and $\alpha_{yz}$ may in principle be nonzero, and the number of field
points required grows to thirteen or to twenty-five, depending on how many
of these off-diagonal components symmetry still forbids.

Atoms provide an additional, essentially free consistency check. In free
space the polarizability of an atom is isotropic, since there is no
preferred direction. Once the cavity is switched on, the polarization axis of the mode
singles out a direction, and the polarizability is in general no
longer isotropic. A nonzero anisotropy $\alpha_{zz}-\alpha_{xx}$ can
appear, reflecting the symmetry breaking imposed by the cavity
itself. Since
this anisotropy must vanish exactly in the absence of coupling, its value
at $\lambda=0$ provides a stringent internal test of the numerical
procedure, independent of any comparison to reference data.

\section{Computational details}
\label{sec:comp}

All calculations use the \texttt{PySCF} library~\cite{Sun2018} for
integrals and for the electronic FCI contraction, while the electron-boson
QED-FCI code that couples this machinery to the photon degrees of freedom
is our own implementation.
For H$_2$ and LiH we use the Sadlej polarized basis
set~\cite{Sadlej1992,Baranowska2009}, which was contracted specifically
for electric response properties and is therefore a natural choice for
computing polarizabilities. Geometries are centered on the center of
nuclear charge, since this removes the origin dependence of the dipole
self-energy for the neutral systems studied here. For LiH we additionally
freeze the lowest molecular orbital. It is well established that the core
polarizability of this molecule is small, so that the valence electrons
account for essentially the entire polarizability response. We tested the resulting error by
comparing to an all-electron calculation at $\lam=0$ and found that it
stays between $-0.22\%$ and $+0.06\%$ over the entire range
$1.0 \le R \le 5.5$~\AA. Because this error is nearly constant across the
whole range, it largely cancels when we later compute the cavity-induced
change $\alpha(\lam)-\alpha(0)$, which is the quantity of physical
interest.

For helium we use the aug-cc-pVTZ basis, augmented by three additional
even-tempered diffuse $s$ and $p$ shells with a ratio of three between
consecutive exponents, giving 35 orbitals in total. This extension is
needed because the metastable excited states of helium are far more diffuse than
the ground state, so that the diffuse tail of the basis, rather than its
description of the atomic core, controls the accuracy of the computed
polarizability.

The photon space is truncated at $N_{\rm p}=8$ for H$_2$, at $N_{\rm
p}=20$ for LiH, and at $N_{\rm p}=4$ for helium. In each case the
truncation was verified by comparing the energies obtained with
successive values of $N_{\rm p}$, and the residual change remained
below $10^{-9}$~a.u.\ in every production run. The much larger value needed for LiH reflects the behavior of the
coherent-state displacement $z$ near dissociation. As the bond is
stretched through the ionic-covalent crossover, the dipole
fluctuations that determine $z$ grow substantially, and a
correspondingly larger photon space is required to converge the
energy at the longest bond lengths sampled, even though far fewer
photon states already suffice near equilibrium.

We performed several additional checks to validate the implementation.
At $\lam=0$ the QED-FCI energy reproduces the energy of a plain electronic
FCI calculation to $4\times10^{-13}$~a.u. The energy is also independent
of the coherent-state displacement $z$ to $2\times10^{-9}$~a.u., as it
should be, since the displacement corresponds to a unitary transformation
of the Hamiltonian and cannot change its spectrum. The frozen core orbitals are removed from the active space, but their
electrons continue to interact with the active electrons. This interaction is incorporated exactly through a standard
frozen-core transformation. The core--active Coulomb and exchange
contributions are absorbed into an effective one-electron operator,
while the core--core repulsion adds a constant energy shift. Crucially, because the dipole
self-energy has already been folded into the one- and two-electron
integrals before this step, the same transformation automatically
captures the core contribution to the self-energy without any additional
terms. We verified this procedure by comparing against a CASCI (Complete Active Space Configuration Interaction)
calculation on the same field-free molecular orbitals, obtaining
agreement to better than $2\times10^{-14}$~a.u.

Free-space polarizabilities provide a further, fully independent check
against previously published results. For helium, the values $\bar\alpha=1.3884$, $316.007$, and
$803.56$~a.u.\ are obtained for the $1\,^1S_0$, $2\,^3S_1$, and
$2\,^1S_0$ states, respectively. These agree with the exact
variational values $1.38319$, $315.6315$, and $800.3163$~a.u.\ of
Refs.~\cite{Yan2000,YanBabb1998} to within $0.08\%$, $0.08\%$, and
$0.41\%$, respectively. For LiH at $R_e$ we obtain
$|d_z|=2.3254$~a.u., within $0.5\%$ of the measured dipole moment. 

\section{Results}
\label{sec:results}

\subsection{Molecular hydrogen}
\label{ssec:H2}

\begin{figure}[!hbtp]
\includegraphics[width=0.9\columnwidth]{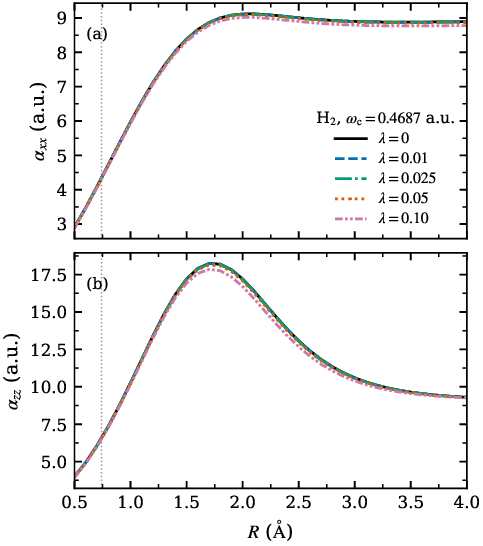}
\caption{Polarizability tensor components of $\mathrm{H}_2$ as a function of the bond length $R$, calculated at the QED-FCI/Sadlej level with cavity frequency $\omega_{\mathrm{cav}} = 0.4687$~a.u.\ and coupling strengths up to $\lambda = 0.1$~a.u. The cavity mode is polarized along the molecular axis ($z$). (a) Perpendicular component $\alpha_{xx}$. (b) Parallel component $\alpha_{zz}$. The vertical dotted line indicates the equilibrium bond length $R_{\mathrm{e}} = 0.7414$~\AA.}
\label{fig:h2abs}%
\end{figure}

The molecular hydrogen (H$_2$) serves as the simplest prototype system, 
where the electronic part of the wave function is treated exactly by both 
QED-FCI and QED-CCSD. Consequently, any discrepancy between the two methods 
can be attributed entirely to the treatment of the photon field. 
Fig.~\ref{fig:h2abs} shows the two independent components of the 
polarizability tensor of $\text{H}_2$ along the bond coordinate for coupling 
strengths up to $\lam = 0.1$~a.u.
On the 
absolute scale, the curves are virtually indistinguishable, and the cavity does 
not qualitatively alter the bond-length dependence. Since H$_2$ is homonuclear, 
it lacks a permanent dipole moment and couples to the cavity field solely 
through the fluctuation of its instantaneous dipole rather than its mean value. 
The resulting cavity effect is therefore weak throughout. 

\begin{figure}[hbtp]
\includegraphics[width=0.9\columnwidth]{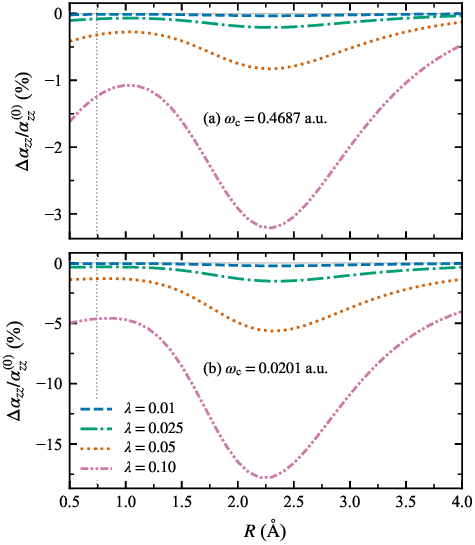}
\caption{Relative change (in percent) of the parallel polarizability of $\mathrm{H}_2$ as a function of the bond length $R$. The cavity mode is polarized along the molecular axis ($z$). (a) Electronic resonance regime with $\omega_{\mathrm{cav}} = 0.4687$~a.u. (b) Vibrational resonance regime with $\omega_{\mathrm{cav}} = 0.02005$~a.u.\ ($\omega_{\mathrm{e}} = 4401~\mathrm{cm}^{-1}$). The vertical dotted line indicates the equilibrium bond length $R_{\mathrm{e}} = 0.7414$~\AA.}
\label{fig:h2rel}
\end{figure}

To illustrate this observation, Fig.~\ref{fig:h2rel} shows the relative change 
of $\alpha_{zz}$ with respect to its free-space value as a function of the 
internuclear distance.
At short bond lengths, the cavity effect is suppressed because the two electrons 
are strongly localized in the symmetric internuclear region. This rigid bond 
restricts electronic mobility and minimizes dipole fluctuations, thereby 
diminishing the light-matter coupling at compressed geometries. As the bond 
is stretched, the polarizability change remains negative and scales with 
the coupling strength, exhibiting a non-monotonic dependence on $R$ with a distinct 
maximum near $R = 2.3$~\AA. This maximum reflects the evolution of the dipole 
fluctuation upon bond elongation: as $R$ increases, the electrons become less 
shared and increasingly localized on individual atoms, boosting the dipole 
variance $\langle (\Delta \hat{d}_z)^2 \rangle$. Beyond this point, the bond 
is essentially broken, the molecule behaves as two isolated atoms, and the 
fluctuation saturates.

The magnitude of the effect also depends strongly on the cavity
frequency. Part of the suppression is undone by the exchange of photons
between the molecule and the cavity, and this exchange weakens as the
cavity frequency is lowered. Photon exchange requires the electrons to
absorb and re-emit a quantum of energy $\wcav$, so the process becomes
less effective when $\wcav$ is small. At $R_e$ and $\lam = 0.05$~a.u. this is reflected in the suppression
of $\alpha_{zz}$, which reaches $0.32\%$ at $\wcav = 0.4687$~a.u.
whereas at $\wcav = 0.0201$~a.u. it grows to $1.31\%$. The same
ordering persists at the maximum near $R\simeq 2.3$~\AA, where the two
values are $0.83\%$ and $5.65\%$, respectively. A
lower cavity frequency therefore always gives the larger effect, since
weaker photon exchange leaves more of the underlying suppression
unopposed, as we show in Section~\ref{ssec:decomp}.

\subsection{Two competing mechanisms}
\label{ssec:decomp}

The frequency dependence has a simple origin. Setting the bilinear term of
Eq.~\eqref{eq:HPF} to zero leaves the dipole self-energy alone. This is the
limit $\wcav \to 0$, because the bilinear term carries the factor
$\sqrt{\wcav/2}$ while the self-energy carries none. The difference between
the full result and this limit is the contribution of photon exchange.

\begin{figure}[t]
\includegraphics[width=0.9\columnwidth]{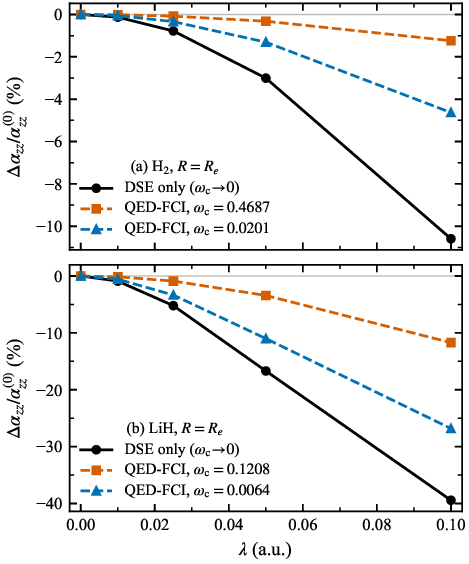}
\caption{Relative change of the parallel polarizability component $\alpha_{zz}$ at the
equilibrium geometry as a function of the coupling strength $\lambda$ for
(a)~$\mathrm{H}_2$ and (b)~$\mathrm{LiH}$. The cavity mode is polarized along
the molecular axis. The black curve with circular markers corresponds to
the limit $\omega_{\mathrm{c}} \to 0$. The orange and blue curves show the
QED-FCI results for the two cavity frequencies considered for each molecule.}
\label{fig:decomp}
\end{figure}

Figure~\ref{fig:decomp} shows the relative change of $\alpha_{zz}$ at the
equilibrium geometry as a function of the coupling strength. In the
$\wcav\to 0$ limit, where only the self-energy remains, the curve is the same
for both cavity frequencies, as it should be. It suppresses $\alpha_{zz}$ by
$10.6\%$ for H$_2$ and by $39.4\%$ for LiH at $\lam = 0.1$~a.u. Photon exchange
acts in the opposite direction and returns $9.4$ and $6.0$ percentage points for
H$_2$ at $\wcav = 0.4687$ and $0.0201$~a.u., and $27.7$ and $12.7$ points for
LiH at $\wcav = 0.1208$ and $0.0064$~a.u., respectively.

The fraction of the self-energy suppression that is cancelled by photon
exchange is $88\%$ for H$_2$ at the higher frequency and $56\%$ at the
lower one. For LiH the corresponding numbers are $70\%$ and $32\%$.
The cancellation between the dipole self-energy and matter-field energy exchange contributions is therefore controlled by the cavity frequency. 
This follows
from second-order perturbation theory in the bilinear term, where the
photon-mediated contribution scales as $\wcav/(\Delta E + \wcav)$ and
vanishes at low frequency.

This decomposition explains a practical difficulty. The observable effect
is a difference of two contributions that are individually much larger. Any
method that treats the two terms with different accuracy will misrepresent
the residue. Mean-field QED-HF contains the self-energy but no photon
exchange, so it overestimates the suppression by a large factor. Truncated
QED coupled cluster contains only part of the photon exchange and lies
between the two limits. Specifically, for comparison we use QED-CCSD-U22-S2, that is, coupled-cluster theory
with single and double electronic excitations extended to the coupled
electron-photon problem. In notation U2$n$-S$m$ the index
$n$ gives the truncation level of the coupled electronic-photonic
excitations and $m$ that of the purely photonic ones, here $n = m =
2$. This part of calculations were performed with the
\texttt{OpenMS} package~\cite{OpenMS}. For H$_2$ at $\lam = 0.1$~a.u.\ and
$\wcav = 0.4687$~a.u.\ we find $-1.24\%$ from QED-FCI, while
QED-CCSD-U22-S2 gives $-3.07\%$ and $\wcav \rightarrow 0 $ limit gives $-10.6\%$.
The truncated coupled cluster result substantially overestimates the
effect, even though the electronic part of its ansatz is exact for two
electrons. This is consistent with the observation
in~\cite{Perspective2025}. As an independent check of our QED-FCI
implementation, we repeated the calculation for H$_2$ in the cc-pVDZ
basis and compared it against the QED-FCI implementation of the eT
program~\cite{eT2020}, using the same photon truncation, $N_{\rm p}=8$,
in both codes. Table~\ref{tab:cc_vs_eT} shows that the two independent codes match to 
within $10^{-9}$~a.u. (nHa) across all coupling strengths, confirming the correctness 
of our electron-photon Hamiltonian implementation.
\begin{table}[htbp]
\caption{Total energy of $\mathrm{H}_2$ at $R_e = 0.7414\text{ \AA}$ and $\omega_{\mathrm{cav}} = 0.4687\text{ a.u.}$, computed using the cc-pVDZ basis set. The cavity mode is polarized along the molecular axis. 
Values are reported for QED-CCSD-U22-S2 (second column), our QED-FCI implementation (third column), and the independent QED-FCI code from the \texttt{eT} program~\cite{eT2020} (fourth column).
In both QED-FCI calculations, the photon space is truncated at $N_{\mathrm{p}} = 8$.
All energies are given in Hartree atomic units.
}
\begin{ruledtabular}
\begin{tabular}{cccc}
$\lambda$ & $E_{\rm CCSD}$ & $E_{\rm FCI}$ & $E_{\rm FCI}^{\rm eT}$ \\
\hline
0.000 & $-1.163\;413\;933$ & $-1.163\;413\;934$ & $-1.163\;413\;934$ \\
0.050 & $-1.161\;917\;621$ & $-1.162\;242\;048$ & $-1.162\;242\;048$ \\
0.100 & $-1.157\;438\;932$ & $-1.158\;739\;088$ & $-1.158\;739\;088$ \\
\end{tabular}
\end{ruledtabular}
\label{tab:cc_vs_eT}
\end{table}

This discrepancy is not a peculiarity of one coupling strength. We
repeated the calculations using QED-HF, QED-CCSD-U22-S2, and QED-FCI for
H$_2$ at $R_e$ over a range of coupling strengths extending to
$\lam = 0.15$~a.u., using the zero-field total energy rather than the
polarizability as the point of comparison, see Table~\ref{tab:cc_vs_fci}. 
\begin{table}[htbp]
\caption{
  Cavity-induced energy shift of H$_2$ at $R=R_e$ and
  $\wcav=0.4687$~a.u., from QED-HF, QED-CCSD-U22-S2 and QED-FCI.
  $\Delta E = E(\lambda)-E(0)$. The last column is the ratio of
  the QED-CCSD shift to the exact QED-FCI shift. The cavity mode is polarized along the molecular axis. Energies in mHa.}
\begin{ruledtabular}
\begin{tabular}{d{1.3}d{2.4}d{2.4}d{2.4}c}
\multicolumn{1}{c}{$\lambda$} & \multicolumn{1}{c}{$\Delta E_{\rm HF}$} & \multicolumn{1}{c}{$\Delta E_{\rm CCSD}$} & \multicolumn{1}{c}{$\Delta E_{\rm FCI}$} & $\Delta E_{\rm CCSD}/\Delta E_{\rm FCI}$ \\
\hline
0.000 & 0.0000 & 0.0000 & 0.0000 & \multicolumn{1}{c}{--} \\
0.025 & 0.6437 & 0.3790 & 0.2977 & 1.273 \\
0.050 & 2.5644 & 1.5122 & 1.1896 & 1.271 \\
0.075 & 5.7328 & 3.3891 & 2.6715 & 1.269 \\
0.100 & 10.1050 & 5.9927 & 4.7371 & 1.265 \\
0.125 & 15.6278 & 9.3016 & 7.3774 & 1.261 \\
0.150 & 22.2442 & 13.2906 & 10.5815 & 1.256 \\
\end{tabular}
\end{ruledtabular}
\label{tab:cc_vs_fci}
\end{table}

\begin{figure}[htbp]
\includegraphics[width=0.9\columnwidth]{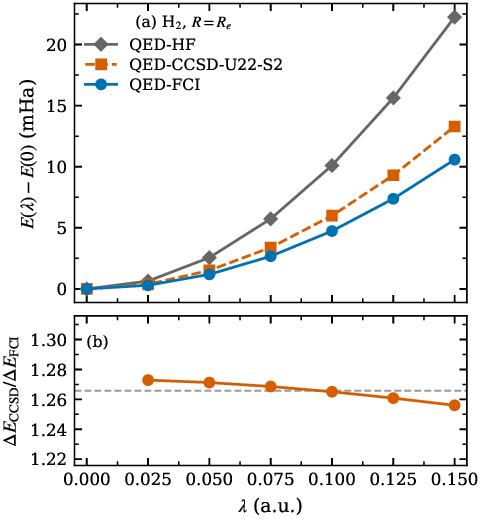}
\caption{Cavity-induced energy shift of H$_2$ at $R=R_e$ and $\wcav=0.4687$~a.u.,
as a function of the coupling strength, from QED-HF, QED-CCSD-U22-S2 and
QED-FCI. The cavity mode is polarized along the molecular axis.
(a) Energy shift $E(\lam)-E(0)$. (b) Ratio of the QED-CCSD to the QED-FCI
shift. The dashed line marks its mean value, 1.266. }
\label{fig:cc_vs_fci}%
\end{figure}

Figure~\ref{fig:cc_vs_fci}
shows the cavity-induced energy shift $\Delta E(\lam) = E(\lam)-E(0)$ for
the three methods, together with the ratio $\Delta E_{\rm
CCSD}/\Delta E_{\rm FCI}$. This ratio is close to $1.27$ at the smallest
coupling shown and decreases only slowly, to $1.26$ at $\lam = 0.15$~a.u.
The truncation error of QED-CCSD accumulates with the
order of the derivative, so that a moderate and roughly constant
overestimate of the energy shift translates into a much larger and
$\lam$-dependent overestimate of the polarizability shift. 
The discrepancy in QED-HF is even more pronounced: the energy-shift ratio to QED-FCI is approximately $2.16$ at $\lam = 0.025$~a.u., whereas at $\lam = 0.15$~a.u. it reduces to $2.10$. This larger ratio, combined with its more pronounced variation across the coupling strengths, confirms that the complete omission of photon exchange represents a far more severe approximation than its truncation at the two-photon level.

The polarizability shift displays a related but quantitatively different trend, see Table~\ref{tab:alpha_cc_vs_fci}.
%\begin{widetext}
\begin{table*}[hbtp]
\caption{
  Isotropic polarizability $\bar{\alpha}$ of H$_2$ at $R=R_e$ and
  $\wcav=0.4687$~a.u., from QED-HF, QED-CCSD-U22-S2 and QED-FCI, in
  atomic units. $\Delta\bar{\alpha}/\bar{\alpha}^{(0)}$ is the relative
  change with respect to the $\lambda=0$ value of each method. The
  last column is the ratio of the QED-CCSD relative change to the
  exact QED-FCI relative change.}
\begin{ruledtabular}
\begin{tabular}{d{1.3}d{1.4}d{1.4}d{1.4}d{2.3}d{2.3}c}
\multicolumn{1}{c}{$\lambda$} & \multicolumn{1}{c}{$\bar\alpha_{\rm HF}$} & \multicolumn{1}{c}{$\bar\alpha_{\rm CCSD}$} & \multicolumn{1}{c}{$\bar\alpha_{\rm FCI}$} & \multicolumn{1}{c}{$\Delta\bar\alpha_{\rm CCSD}/\bar\alpha^{(0)}$ (\%)} & \multicolumn{1}{c}{$\Delta\bar\alpha_{\rm FCI}/\bar\alpha^{(0)}$ (\%)} & ratio \\
\hline
0.000 & 5.1171 & 5.0875 & 5.0875 & 0.000 & 0.000 & \multicolumn{1}{c}{--} \\
0.025 & 5.1026 & 5.0799 & 5.0845 & -0.149 & -0.059 & 2.508 \\
0.050 & 5.0609 & 5.0577 & 5.0755 & -0.586 & -0.237 & 2.477 \\
0.075 & 4.9966 & 5.0222 & 5.0606 & -1.284 & -0.528 & 2.431 \\
0.100 & 4.9160 & 4.9754 & 5.0403 & -2.203 & -0.928 & 2.374 \\
0.125 & 4.8249 & 4.9197 & 5.0148 & -3.300 & -1.429 & 2.310 \\
0.150 & 4.7286 & 4.8571 & 4.9848 & -4.530 & -2.020 & 2.242 \\
\end{tabular}
\end{ruledtabular}
\label{tab:alpha_cc_vs_fci}
\end{table*}
%\end{widetext}

In particular, Table~\ref{tab:alpha_cc_vs_fci} lists the isotropic polarizability 
$\bar{\alpha}=(\alpha_{xx}+\alpha_{yy}+\alpha_{zz})/3$ obtained from these three 
methods across the same set of coupling strengths, along with the ratio of 
the cavity-induced shifts $\Delta\bar\alpha_{\rm CCSD}/\Delta\bar\alpha_{\rm FCI}$. 
This ratio is close to $2.5$ at the weakest coupling shown, consistent with the 
trend found above for the energy shift. However, unlike the 
nearly constant energy-shift ratio, it decreases steadily with $\lam$, 
reaching $2.24$ at $\lam=0.15$~a.u. This distinct behavior arises because the 
energy-shift ratio is primarily determined by the missing electron-photon 
correlation at a given coupling strength. In contrast, the polarizabilities 
ratio additionally reflects how this missing correlation distorts the curvature 
of $E(\lam)$, which softens relative to the exact result as $\lam$ grows. 
Both observations point to the same conclusion: truncated QED-CCSD is not 
a uniformly biased approximation. Its error depends both on the specific 
quantity computed and on the coupling strength. Therefore, its applicability 
should be evaluated with caution beyond the regimes validated against 
an exact reference.

Stretching a chemical bond can induce structural shortcomings that go beyond mere quantitative errors. As the bond stretches toward the dissociation limit, a single Slater determinant becomes qualitatively inadequate to describe the true wave function. Methods such as QED-CCSD naturally inherit this breakdown, as they are fundamentally rooted in a single-determinant reference. For LiH, this limitation of restricted Hartree--Fock is well established even without a cavity, and it is directly confirmed here. As a diagnostic, $\max|T_1|$, the largest single-excitation amplitude of the coupled-cluster wave function, measures how strongly the determinant reference is perturbed by correlation. This amplitude stays below $0.15$ up to $R \simeq 2.8$~\AA, comfortably beyond the equilibrium bond length considered in the preceding sections, but grows to $0.61$ by the stretched bond length $R = 5.0$~\AA. An amplitude of this magnitude signals that the wave function has acquired a pronounced 
multi-determinant character, so that any single-reference treatment ceases to be reliable at this bond length.

The presence of the cavity further exacerbates these computational difficulties. 
At $\lam=0$, the coupled-cluster calculation still reaches numerical convergence 
despite the poor reference description, whereas at $\lam=0.05$ and $\lam=0.10$~a.u., 
the amplitude equations fail to converge altogether, causing the calculation 
to break down. Conversely, QED-FCI does not rely on a single-determinant reference 
and converges cleanly across all coupling strengths studied, yielding cavity-induced 
energy shifts of $7.26$~mHa at $\lam=0.05$~a.u. and $25.73$~mHa at $\lam=0.10$~a.u. 
The practical implication is that single-reference QED-CCSD directly inherits 
the well-known failure of restricted Hartree--Fock at bond dissociation. 
Furthermore, the truncation of the photon sector compounds this issue once 
the cavity is switched on, as the amplitude equations should simultaneously 
compensate for an inadequate electronic reference and account for the missing 
electron-photon correlation. QED-FCI, which is independent of both the choice 
of a reference determinant and photon-number truncation, remains robust throughout 
the entire dissociation curve and provides the necessary reference data in this regime.

\subsection{Lithium hydride}
\label{ssec:LiH}

\begin{figure}[htbp]
\includegraphics[width=0.9\columnwidth]{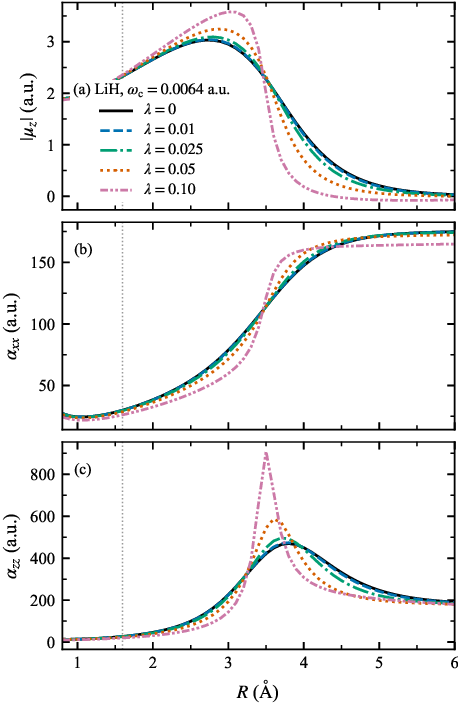}
\caption{\label{fig:lihprops}%
Dipole moment and polarizability of LiH from QED-FCI with a frozen core. (a) Magnitude of the dipole moment. (b) Perpendicular
polarizability. (c) Parallel polarizability. The cavity frequency is
$\wcav = 0.0064$~a.u. The dotted line marks $R_e = 1.5957$~\AA.}
\end{figure}

Unlike the homonuclear H$_2$ molecule discussed above, LiH possesses a large 
permanent dipole moment. Consequently, the dipole self-energy shift is 
already nonvanishing at the reference level, while the polarizability itself 
is larger by an order of magnitude. Figure~\ref{fig:lihprops} displays the 
calculated dipole moment along with the two independent components of the 
polarizability tensor.

\begin{figure}[htbp]
\includegraphics[width=0.9\columnwidth]{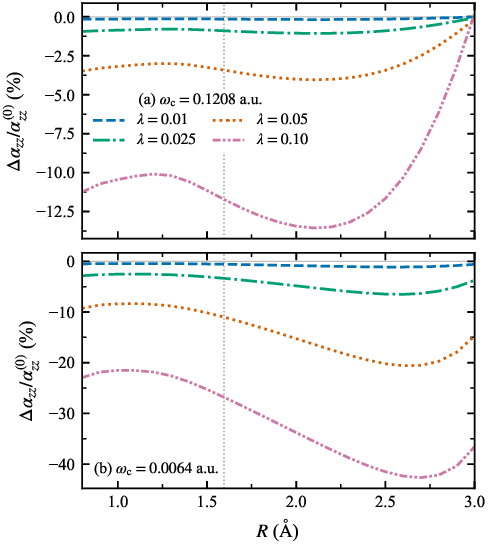}
\caption{\label{fig:lihrel}%
Relative change of the parallel polarizability of LiH. (a) Cavity frequency
near the first electronic excitation. (b) Cavity frequency near the
vibrational frequency $\omega_e = 1405$~cm$^{-1}$.}
\end{figure}

As seen from Fig.~\ref{fig:lihprops}, the dipole moment increases almost 
linearly through the equilibrium region, reaches a maximum near $R = 3.1$~\AA, 
and subsequently drops as the ionic configuration gives way to neutral fragments. 
This behavior reflects the well-known avoided crossing between the ionic 
Li$^+$H$^-$ and covalent configurations. Near equilibrium, the effective charge 
transferred from Li to H is close to one electron, causing the dipole to grow 
linearly with the bond length. Beyond the crossing region, however, this charge 
drops rapidly toward zero, so that the dipole moment first peaks and then vanishes 
in the separated-atom limit. The polarizability exhibits a similar trend, 
displaying a pronounced maximum at comparable bond lengths. The maximum of $\alpha_{zz}$ coincides with the steepest decrease of the
dipole moment, where the $X^1\Sigma^+$ and $A^1\Sigma^+$ states approach
each other and the response is dominated by a single 
transition. The cavity sharpens this avoided crossing, so that the
maximum grows and narrows with increasing coupling strength.
This larger and more polarizable valence electron density also substantially 
enhances the cavity effect compared to that in H$_2$. Figure~\ref{fig:lihrel} 
displays the resulting relative change in $\alpha_{zz}$.

At the equilibrium geometry and $\lam=0.05$~a.u., the suppression amounts 
to $3.4\%$ at the electronic cavity frequency and $11.0\%$ at the vibrational 
frequency, which is approximately an order of magnitude larger than the effect 
observed for H$_2$ at the same coupling strength. The plots also reveal a nonlinear dependence of the molecular response on both the internuclear distance and the cavity coupling strength.

The permanent dipole moment itself, by contrast, is far less sensitive to
the cavity. At $\lam=0.1$~a.u.\ and $\wcav=0.0064$~a.u.\ it changes by
only $1.3\%$ at $R_e$, compared with $26.8\%$ for $\alpha_{zz}$. This
difference reflects the distinct nature of the two quantities. The cavity
modifies how the electron density responds to a field, rather than the
equilibrium density itself, and the polarizability is far more sensitive
to this change than the static dipole moment. Table~\ref{tab:mol} collects
the equilibrium values for both molecules.
\begin{widetext}

\begin{table}[htbp]
\caption{%
  Static polarizability (in atomic units) of H$_2$ at $R_e=0.7414$~\AA{} and 
of LiH at $R_e=1.5957$~\AA\, from QED-FCI/Sadlej calculations. The cavity mode 
is polarized along the molecular axis. The column $\wcav\to 0$ represents 
the dipole self-energy limit and is independent of the cavity frequency. 
Relative changes are evaluated with respect to the $\lambda=0$ values.
}
  \label{tab:mol}
\begin{ruledtabular}
\begin{tabular}{lcd{2.4}d{2.4}d{2.4}d{3.3}d{3.3}}
System & $\lambda$ (a.u.) & \multicolumn{1}{c}{$\alpha_{zz}$} & \multicolumn{1}{c}{$\alpha_{xx}$} & \multicolumn{1}{c}{$\alpha_{zz}(\omega_{\rm c}\!\to\!0)$} & \multicolumn{1}{c}{$\Delta\alpha_{zz}/\alpha_{zz}^{(0)}$ (\%)} & \multicolumn{1}{c}{$\Delta\alpha_{xx}/\alpha_{xx}^{(0)}$ (\%)} \\
\hline
H$_2$, free space & 0 & 6.5706 & 4.3458 & \multicolumn{1}{c}{--} & 0.000 & 0.000 \\
H$_2$, $\omega_{\rm c}=0.4687$ & 0.010 & 6.5698 & 4.3455 & 6.5624 & -0.013 & -0.007 \\
 & 0.025 & 6.5654 & 4.3439 & 6.5195 & -0.080 & -0.044 \\
 & 0.050 & 6.5497 & 4.3382 & 6.3731 & -0.319 & -0.175 \\
 & 0.100 & 6.4892 & 4.3157 & 5.8747 & -1.240 & -0.693 \\
H$_2$, $\omega_{\rm c}=0.0201$ & 0.010 & 6.5671 & 4.3447 & 6.5624 & -0.054 & -0.026 \\
 & 0.025 & 6.5485 & 4.3387 & 6.5195 & -0.338 & -0.163 \\
 & 0.050 & 6.4847 & 4.3180 & 6.3731 & -1.308 & -0.640 \\
 & 0.100 & 6.2663 & 4.2416 & 5.8747 & -4.631 & -2.397 \\
 \hline
LiH, free space & 0 & 26.5989 & 29.9508 & \multicolumn{1}{c}{--} & 0.000 & 0.000 \\
LiH, $\omega_{\rm c}=0.1208$ & 0.010 & 26.5603 & 29.9297 & 26.3590 & -0.145 & -0.070 \\
 & 0.025 & 26.3606 & 29.8194 & 25.2114 & -0.896 & -0.439 \\
 & 0.050 & 25.6881 & 29.4338 & 22.1564 & -3.424 & -1.726 \\
 & 0.100 & 23.4812 & 28.0332 & 16.1094 & -11.721 & -6.402 \\
LiH, $\omega_{\rm c}=0.0064$ & 0.010 & 26.4465 & 29.8915 & 26.3590 & -0.573 & -0.198 \\
 & 0.025 & 25.7071 & 29.5927 & 25.2100 & -3.353 & -1.196 \\
 & 0.050 & 23.6716 & 28.6647 & 22.1463 & -11.005 & -4.294 \\
 & 0.100 & 19.4650 & 26.1483 & 16.0897 & -26.820 & -12.696 \\
\end{tabular}
\end{ruledtabular}
\end{table}
\end{widetext}

\subsection{Isotropic response and mode orientation}
\label{ssec:orient}

\begin{figure}[htbp]
\includegraphics[width=0.9\columnwidth]{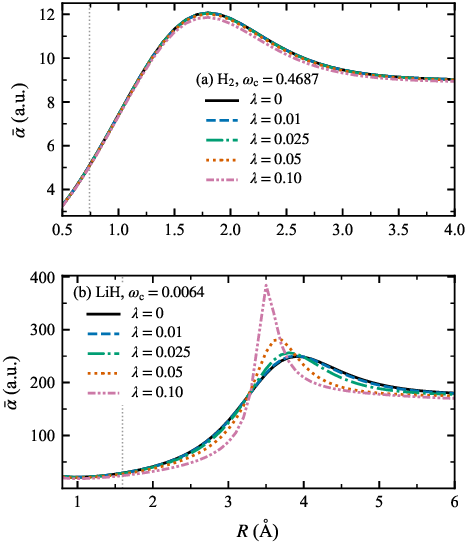}
\caption{%
Isotropic polarizability $\bar{\alpha}=(\alpha_{xx}+\alpha_{yy}+\alpha_{zz})/3$
as a function of the bond length, for the coupling strengths of
Fig.~\ref{fig:h2abs}. (a) H$_2$, $\wcav=0.4687$~a.u. (b) LiH,
$\wcav=0.0064$~a.u. The dotted vertical line marks the equilibrium bond
length.}
\label{fig:isotropic}
\end{figure}

While the results above focus on the parallel component $\alpha_{zz}$, an experiment that does not resolve the tensor depends on the isotropic average $\bar{\alpha} = \tfrac{1}{3}(\alpha_{xx}+\alpha_{yy}+\alpha_{zz})$. The corresponding curves for both molecules are displayed in Fig.~\ref{fig:isotropic}.
The cavity-induced changes are discernible but small on this absolute scale,
consistent with the relative changes reported above when the perpendicular
components are included in the averaging. The $\mathrm{H}_2$ molecule exhibits
the weakest response to the cavity, consistent with its lack of a permanent
dipole moment. 

\begin{figure*}[!htbp]
\includegraphics[width=\textwidth]{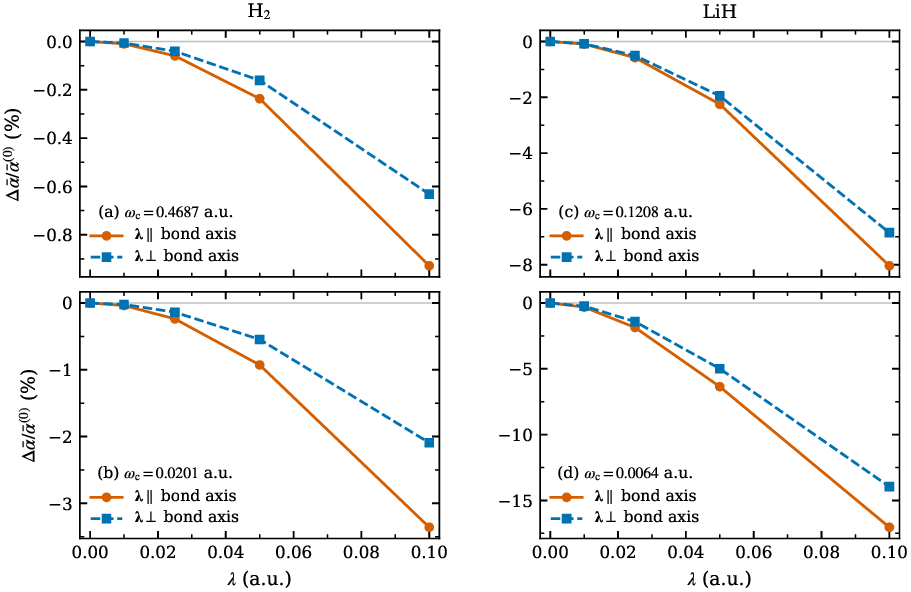}
\caption{%
Relative change of the isotropic polarizability at $R=R_e$, as a function
of the coupling strength, for the cavity mode polarized parallel and
perpendicular to the bond axis. (a,b) H$_2$, at the electronic and
vibrational cavity frequency. (c,d) LiH, at the electronic and vibrational
cavity frequency.}
\label{fig:orient}
\end{figure*}

All results discussed thus far employ a cavity mode polarized along the 
molecular axis. While this choice is not unique, it is important to examine 
the extent to which the cavity effect persists when the mode is polarized 
perpendicularly to the bond. To isolate the role of the mode orientation from 
that of the coupling strength itself, we computed both configurations at 
identical coupling strengths for both molecules. Figure~\ref{fig:orient} 
compares the resulting relative changes in $\bar{\alpha}$ for the two orientations.
%\begin{widetext}
%\end{widetext}

The cavity
suppresses the isotropic polarizability in every case, so the sign of the
effect does not depend on the mode direction, but the parallel orientation
is consistently more effective than the perpendicular one. This ordering
follows from Eq.~\eqref{eq:HCS}. The dipole fluctuation that enters the
self-energy is largest along the bond axis, where the electron density is
most polarizable, so a mode aligned with this axis samples the largest
fluctuation and produces the strongest suppression.

The size of the anisotropy between the two orientations differs markedly
between the two molecules. For H$_2$ at $R_e$ and $\lam = 0.1$~a.u. the
suppression is larger by a factor of $1.5$ to $1.7$ for the parallel
orientation, depending on the cavity frequency. For LiH the same
comparison gives a factor of only $1.2$ to $1.3$. The weaker anisotropy of
LiH is consistent with its larger permanent dipole moment, which makes the
electron density respond more uniformly to a field applied along any
direction and reduces the directional preference of the dipole
fluctuation that Eq.~\eqref{eq:HCS} singles out.

\subsection{Helium clock states}
\label{ssec:He}

Investigating how a cavity affects the systems utilized in atomic clocks is 
of immediate relevance. The two molecules studied above already provide important 
insights into one broad class of such devices. Many proposed molecular clocks, 
such as those based on N$_2^+$ and O$_2^+$~\cite{Kajita2014}, employ a vibrational 
transition within the ground electronic state. The frequency of such a transition 
is shifted by the blackbody radiation field in proportion to the fourth power 
of the temperature and to the static polarizabilities of the two vibrational 
levels involved~\cite{Zalialiutdinov2024}. This thermal shift currently 
constitutes one of the dominant contributions to the systematic uncertainty budget. 
Given that the cavity suppresses the ground-state polarizability of H$_2$ and LiH, 
it should correspondingly screen this class of clocks from a portion of their 
primary systematic shift. In contrast, optical clocks often rely on a fundamentally 
different type of transition, which is strongly forbidden and connects excited 
electronic states rather than vibrational levels of the ground state. The clock 
transition of helium considered here represents the simplest example of this second 
class~\cite{Rooij2011,Notermans2014}, and it remains to be seen whether the 
suppression observed for ground-state molecules extends to such electronic transitions.

The forbidden transition $2\,^3S_1 \to 2\,^1S_0$ of helium at 1557~nm 
has been measured with a fractional accuracy of $10^{-12}$~\cite{Rooij2011,Notermans2014}. 
Its blackbody radiation shift is governed by the difference between the 
polarizabilities of these two states. Furthermore, helium serves as an ideal 
prototype system for QED-FCI calculations. Since the electronic part of the 
problem is treated exactly for a two-electron system, any cavity-induced 
modifications cannot be attributed to an approximate treatment of electron correlation.

\begin{figure}[htbp]
\centering
\includegraphics[width=0.9\columnwidth]{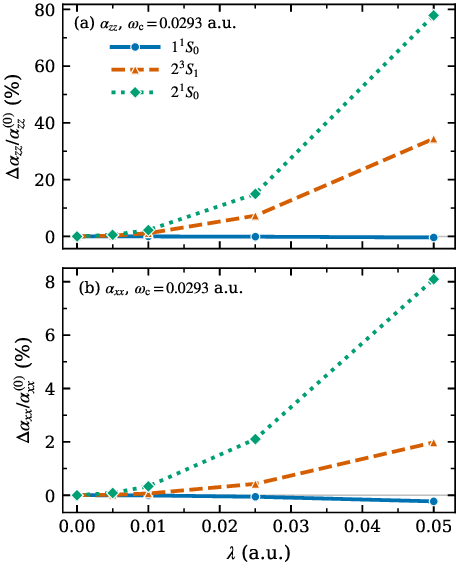}
\caption{\label{fig:hestates}%
Relative change of the polarizability of the three lowest $S$ states of
helium as a function of the coupling strength, from QED-FCI in the
augmented aug-cc-pVTZ basis, for a cavity tuned to the clock transition,
$\wcav=0.0293$~a.u.
Solid lines with circles, dashed lines with triangles, and dotted lines
with diamonds denote $1\,^1S_0$, $2\,^3S_1$, and $2\,^1S_0$, respectively.
(a) Parallel component, $\alpha_{zz}$.
(b) Perpendicular component, $\alpha_{xx}$.}
\end{figure}

The free-space polarizabilities of the three lowest $S$ states differ by 
almost three orders of magnitude. Within the present approach, we obtained 
$1.3884$, $316.007$, and $803.56$~a.u. for the $1\,^1S_0$, $2\,^3S_1$, and 
$2\,^1S_0$ states, respectively. These values are in excellent agreement with the 
variational results reported in~\cite{Yan2000}. The polarizability difference for the clock transition is $487.55$~a.u.,
the largest value among the systems considered in this work.
Its cavity-induced modification directly controls the corresponding
blackbody radiation shift. Figure~\ref{fig:hestates} shows how the coupling to the cavity changes
each of the three polarizabilities. 

The ground state remains virtually insensitive, exhibiting a change below 
$0.3\%$ at $\lam=0.05$~a.u. In contrast, the two metastable states respond 
strongly, showing a behavior opposite to that observed for the molecules: 
their polarizabilities increase rather than decrease. Specifically, at 
$\wcav=0.0293$~a.u.\ and $\lam=0.05$~a.u., the parallel component increases 
by $34\%$ for the $2\,^3S_1$ state and by $78\%$ for the $2\,^1S_0$ state.

This sign reversal relative to the molecular case (see Fig.~\ref{fig:orient}) traces back to the small energy 
denominator $E_{2P}-E_{2S}$ entering the sum-over-states expression for the 
polarizability. For the metastable states, the nearest $P$ state lies 
only $0.022$~a.u. away. Since the dipole fluctuation of the $S$ state is 
larger, the self-energy pushes its energy level upward more strongly than 
that of the $P$ state. The effective gap between them is therefore reduced, 
causing the polarizability to grow instead of diminish. For the molecular ground 
states, the energy gap to the first excited state is sufficiently large that 
this shifting mechanism is absent, and the stiffening of the electron density 
dominates instead.

The response anisotropy, defined as the geometric difference between the polarizability components along and perpendicular to the cavity axis, confirms that this growth is entirely cavity-induced. This aligns with the spherical symmetry of a free atom, which renders its polarizability tensor isotropic. In our calculations, this anisotropy is below $10^{-4}$~a.u.\ at $\lam=0$ for all three states, becoming appreciable only under light-matter coupling. Specifically, at $\lam=0.05$~a.u. and the near-resonant frequency $\wcav$, it reaches $103$~a.u. for the $2\,^3S_1$ state and $561$~a.u. for the $2\,^1S_0$ state.

\begin{figure}[h!]
\includegraphics[width=\columnwidth]{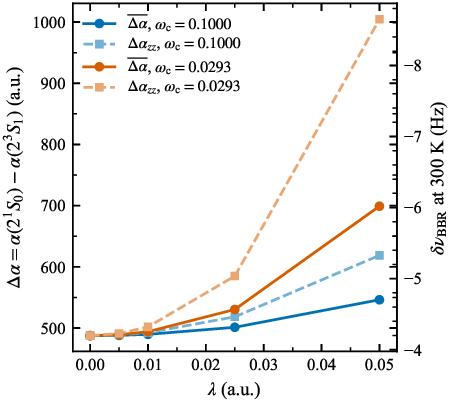}
\caption{\label{fig:hediff}%
Polarizability difference $\Delta\alpha=\alpha(2\,^1S_0)-\alpha(2\,^3S_1)$ 
of the helium clock transition vs coupling strength. Solid lines with circles 
and dashed lines with squares represent the isotropic average and parallel component, 
respectively. The right axis indicates the corresponding blackbody radiation 
shift at 300~K, obtained from $\delta\nu=-\tfrac12\langle E^2\rangle\overline{\Delta\alpha}$ \cite{Farley1981}.
}
\end{figure}

Since the metastable states exhibit a substantially stronger response than 
the ground state, the differential polarizability of the clock transition itself 
increases significantly inside the cavity. Figure~\ref{fig:hediff} displays 
this growth as a function of the coupling strength for both cavity frequencies 
under consideration. 

At $\lam=0.05$~a.u., the isotropic value increases from 
the free-space value of $487.55$~a.u. to $546.21$~a.u. at $\wcav=0.1$~a.u. 
and to $699.07$~a.u. at $\wcav=0.0293$~a.u., corresponding to relative changes 
of $12\%$ and $43\%$, respectively.

This growth translates directly into a larger blackbody radiation shift through $\delta\nu = -\tfrac{1}{2}\langle E^2\rangle\,\overline{\Delta\alpha}$, where $\langle E^2\rangle = (831.9~{\rm V/m})^2 (T/300~{\rm K})^4$ \cite{Farley1981,Itano1982}. In free space, we obtain a shift of $-4.198$~Hz at 300~K, corresponding to a 
fractional shift of $2.18\times10^{-14}$ of the transition frequency. This value agrees to within 5\% with the dynamic calculation made in~\cite{Zalialiutdinov2017He}, which gives $-4.398$~Hz. The difference
arises because blackbody radiation at 300~K is not a static field. Its
typical frequency, about 0.004~a.u., is small but not negligible compared
with the 0.022~a.u.\ gap between the $2\,^1S$ and $2\,^1P$ states, which
governs the response of the $2\,^1S$ state. Since each term of the sum over
states grows with frequency as $1/(\Delta E^2-\omega^2)$, the thermal field
probes a response about 3.5\% stronger than the static one for $2\,^1S$ and
about 1\% stronger for $2\,^3S_1$, whose nearest $P$ state lies twice as
far away. With this leading dynamic correction the static FCI result becomes
$-4.41$~Hz, within 0.3\% of the fully dynamic value. This agreement with
an independent variational calculation provides an additional check of the
free-space polarizabilities obtained here.

Introducing the cavity further increases the magnitude of the blackbody
shift. At $\lam=0.05$~a.u., it reaches $-4.703$~Hz for the off-resonant
cavity and $-6.019$~Hz for the resonant one, corresponding to fractional
shifts of $2.44\times10^{-14}$ and $3.13\times10^{-14}$, respectively. This behavior is the opposite of our findings for considered molecular system. In the present atomic case, the cavity enhances the polarizability 
of the metastable states, making the clock transition more sensitive to the thermal field. Crucially, the resulting shift remains an order of magnitude below the current experimental accuracy of the transition. This insignificance opens up feasible opportunities for further high-precision clock measurements directly inside cavity environments. 

Table~\ref{tab:he} lists the computed blackbody radiation shifts together with the underlying polarizabilities.
\begin{widetext}

\begin{table*}[h!]
\caption{Static polarizabilities of the $1^1S_0$, $2^3S_1$, and $2^1S_0$ states of helium in a cavity, computed using QED-FCI with the aug-cc-pVTZ basis set augmented by three even-tempered diffuse $s$ and $p$ shells.
Here, $\overline{\Delta\alpha}$ denotes the differential polarizability of the $2^3S_1 \to 2^1S_0$ clock transition, while $\delta\nu_{\mathrm{BBR}}$ represents the corresponding blackbody radiation shift at $300\text{ K}$.
The last column lists the fractional shift relative to the transition frequency.
All polarizabilities are reported in atomic units.}
\begin{ruledtabular}
\begin{tabular}{cd{1.3}d{1.4}d{3.3}d{4.2}d{3.2}d{2.4}c}
$\omega_{\rm c}$ (a.u.) & \multicolumn{1}{c}{$\lambda$} & \multicolumn{1}{c}{$\bar\alpha(1^1S_0)$} & \multicolumn{1}{c}{$\bar\alpha(2^3S_1)$} & \multicolumn{1}{c}{$\bar\alpha(2^1S_0)$} & \multicolumn{1}{c}{$\overline{\Delta\alpha}$} & \multicolumn{1}{c}{$\delta\nu_{\rm BBR}$ (Hz)} & $\delta\nu_{\rm BBR}/\nu$ \\
\hline
0.1000 & 0.000 & 1.3884 & 316.007 & 803.56 & 487.55 & -4.1980 & $2.18\times10^{-14}$ \\
 & 0.005 & 1.3883 & 316.103 & 804.19 & 488.08 & -4.2025 & $2.18\times10^{-14}$ \\
 & 0.010 & 1.3882 & 316.392 & 806.08 & 489.69 & -4.2164 & $2.19\times10^{-14}$ \\
 & 0.025 & 1.3875 & 318.482 & 819.69 & 501.20 & -4.3155 & $2.24\times10^{-14}$ \\
 & 0.050 & 1.3850 & 326.783 & 872.99 & 546.21 & -4.7030 & $2.44\times10^{-14}$ \\
0.0293 & 0.000 & 1.3884 & 316.007 & 803.56 & 487.55 & -4.1980 & $2.18\times10^{-14}$ \\
 & 0.005 & 1.3883 & 316.331 & 805.48 & 489.15 & -4.2117 & $2.19\times10^{-14}$ \\
 & 0.010 & 1.3882 & 317.315 & 811.32 & 494.01 & -4.2535 & $2.21\times10^{-14}$ \\
 & 0.025 & 1.3874 & 324.592 & 854.97 & 530.37 & -4.5667 & $2.37\times10^{-14}$ \\
 & 0.050 & 1.3844 & 356.452 & 1055.52 & 699.07 & -6.0192 & $3.13\times10^{-14}$ \\
\end{tabular}
\end{ruledtabular}
\label{tab:he}
\end{table*}
\end{widetext}

\section{Conclusions}
\label{sec:concl}

For the systems considered in this work, an optical cavity does not act on the static polarizability as a uniform perturbation. Instead, it modifies the molecular response through two competing mechanisms of opposite sign. The dipole self-energy stiffens the electron density, thereby suppressing the response, whereas the photon-exchange contribution acts in the opposite direction and scales in magnitude with the cavity frequency. For the molecular 
ground states of H$_2$ and LiH, these two contributions nearly cancel each other, leaving only a small residual suppression of the polarizability. To our knowledge, the dependence of this response
on the cavity coupling strength has not previously been systematically
examined at different internuclear separations. This near cancellation makes the polarizability a highly demanding test of electron-photon correlation. Mean-field theory and truncated coupled-cluster theory fail to reproduce this behavior at large interatomic separations. This failure arises because both methods restrict the electron-photon wave function to a small number of photon excitations, a limitation that becomes increasingly inadequate 
as the bond stretches toward dissociation.

The near cancellation of these two mechanisms has direct practical implications. A cavity that suppresses the ground-state polarizability of a molecule simultaneously reduces its sensitivity to slowly varying external fields, including the thermal electromagnetic field of the environment. Therefore, it can shield the molecule from the blackbody radiation shift that limits the accuracy of vibrational frequency standards, such as those proposed for N$_2^+$ and O$_2^+$. This suppression should also reduce the rate of blackbody-induced photo-dissociation, which currently restricts the trapping lifetime of such molecular ions. More generally, it should modify van der Waals interactions, since these are governed by the dynamic polarizability, of which the static response computed here represents the zero-frequency limit. Given that the cavity suppresses the ground-state polarizability of
H$_2$ and LiH, it should correspondingly screen this class of clocks from a
portion of their primary systematic shift.

This expectation can be checked directly for the simplest molecular clock
candidate, the one-electron ion H$_2^+$~\cite{Schiller2014}, for which the
electron-photon problem can be solved exactly. The calculation was
performed in a Gaussian basis set, with clamped nuclei at $R=2.0$~bohr and
the cavity tuned to the $v=0\to1$ vibrational transition,
$\wcav=0.00998$~a.u. In free space we obtain $\bar\alpha=2.8641$~a.u., in
agreement with the exact value of 2.8643~a.u.~\cite{BishopCheung1978}. For
couplings up to $\lam=0.1$~a.u. the isotropic polarizability changes by
less than 0.1\% when the mode is polarized along the bond and decreases by
about 1\% for a perpendicular mode. For H$_2^+$ the screening is thus
present but weak, and its size is set mainly by the orientation of the
mode.

Whether such changes matter in practice depends on how strong the coupling
can be made. Vacuum-induced shifts of clock transitions are already
resolved in current experiments, for example the millihertz cooperative
Lamb shifts in a strontium lattice clock~\cite{Hutson2024} and the
Casimir-Polder shift of atoms near a surface~\cite{CP2026}. The coupling
strengths considered here correspond to mode volumes of 0.19--19~nm$^3$,
which are within reach of plasmonic nanocavities~\cite{Chikkaraddy2016}. In
such environments the cavity-induced change of the polarizability becomes
a sizable effect, whereas in the much larger volumes of conventional traps
it is negligible, since it scales as $\lam^2$.

Existing strategies for suppressing the blackbody shift rely either on combining two clock transitions into a synthetic frequency largely immune to thermal perturbations~\cite{Yudin2011} or on cryogenic operation. A passive, vacuum-induced screening of the kind identified here offers an alternative route to this goal. Crucially, this mechanism applies already at weak coupling, well below the regime where the molecule is appreciably dressed by the cavity mode and where its coherence is limited by the photon lifetime, as is typically the case for strongly coupled polaritons~\cite{OpenMS}. Cavity-based suppression of blackbody-induced transitions has already
been demonstrated for circular Rydberg atoms at room
temperature~\cite{Holzl2024,Pultinevicius2026}.
These advances motivate extending the present analysis to
cavity-induced modifications of thermal shifts in Rydberg states.

The sign of the cavity-induced correction is not universal, however. For the metastable states of helium, the polarizability is dominated by a nearby $P$ state. Since the dipole fluctuation of the $S$ state is larger, the dipole self-energy shifts its energy level upward more strongly than that of the $P$ state. Consequently, the effective energy gap is reduced rather than widened, causing the polarizability to increase. The blackbody 
radiation shift of the $2^3S_1\to 2^1S_0$ clock transition grows accordingly. Depending on the electronic structure, cavity confinement can thus either protect a frequency standard from thermal perturbations or amplify them. Determining the active regime in a given system requires the correlated, non-perturbative treatment developed in this work.

\section*{Data and Software Availability}
\label{sec:data}
The QED-FCI code, along with the input files, output data, and plotting
scripts used in this work, is available at
\url{https://github.com/Physfock/qed-fci}. A snapshot of the repository
has been archived on Zenodo and is available at
\href{https://doi.org/10.5281/zenodo.22798177}{10.5281/zenodo.22798177}.

\section*{Acknowledgments}
The research was supported by the Russian Science Foundation (grant no. 26-72-10004).

\bibliography{refs_fixed}

@article{Pultinevicius2026,
  author  = {Pultinevicius, Einius and
             G{\"o}tzelmann, Aaron and
             Thielemann, Fabian and
             H{\"o}lzl, Christian and
             Meinert, Florian},
  title   = {Long-lived giant circular {Rydberg} atoms at room temperature},
  journal = {Nature Communications},
  volume  = {17},
  pages   = {9834},
  year    = {2026},
  doi     = {10.1038/s41467-026-77764-x},
  url     = {https://doi.org/10.1038/s41467-026-77764-x}
}

@article{Zalialiutdinov2017He,
  title = {BBR-induced Stark shifts and level broadening in a helium atom},
  volume = {51},
  ISSN = {1361-6455},
  url = {http://dx.doi.org/10.1088/1361-6455/aa8e6e},
  DOI = {10.1088/1361-6455/aa8e6e},
  number = {1},
  journal = {Journal of Physics B: Atomic,  Molecular and Optical Physics},
  publisher = {IOP Publishing},
  author = {Zalialiutdinov,  T and Solovyev,  D and Labzowsky,  L},
  year = {2018},
  month = Nov,
  pages = {015003}
}

@article{Holzl2024,
  author  = {H{\"o}lzl, Christian and
             G{\"o}tzelmann, Aaron and
             Pultinevicius, Einius and
             Wirth, M. and
             Meinert, Florian},
  title   = {Long-Lived Circular {Rydberg} Qubits of Alkaline-Earth
             Atoms in Optical Tweezers},
  journal = {Physical Review X},
  volume  = {14},
  pages   = {021024},
  year    = {2024},
  doi     = {10.1103/PhysRevX.14.021024},
  url     = {https://doi.org/10.1103/PhysRevX.14.021024}
}

@article{Kajita2014,
  author  = {Kajita, M. and Gopakumar, G. and Abe, M. and Hada, M. and Keller, M.},
  title   = {Test of $m_p/m_e$ changes using vibrational transitions in {N}$_2^+$},
  journal = {Phys. Rev. A},
  volume  = {89},
  pages   = {032509},
  year    = {2014},
  doi     = {10.1103/PhysRevA.89.032509}
}

@article{Yudin2011,
  author  = {Yudin, V. I. and Taichenachev, A. V. and Okhapkin, M. V. and
            Bagayev, S. N. and Tamm, Chr. and Peik, E. and
            Huntemann, N. and Mehlst{\"a}ubler, T. E. and Riehle, F.},
  title   = {Atomic clocks with suppressed blackbody radiation shift},
  journal = {Phys. Rev. Lett.},
  volume  = {107},
  pages   = {030801},
  year    = {2011},
  doi     = {10.1103/PhysRevLett.107.030801}
}

@article{Zanon2006,
  author  = {Zanon-Willette, T. and Ludlow, A. D. and Blatt, S. and Boyd, M. M.
              and Arimondo, E. and Ye, J.},
  title   = {Cancellation of {S}tark Shifts in Optical Lattice Clocks by Use of
             Pulsed Raman and Electromagnetically Induced Transparency Techniques},
  journal = {Phys. Rev. Lett.},
  volume  = {97},
  pages   = {233001},
  year    = {2006},
  doi     = {10.1103/PhysRevLett.97.233001}
}

@article{Zalialiutdinov2024,
  title = {Blackbody-radiation vibrational level shifts in the ground electronic state of {N}$_{2}^{+}$},
  author = {Zalialiutdinov, T. and Demidov, Y. and Solovyev, D.},
  journal = {Phys. Rev. A},
  volume = {110},
  issue = {6},
  pages = {063123},
  numpages = {8},
  year = {2024},
  month = {Dec},
  publisher = {American Physical Society},
  doi = {10.1103/PhysRevA.110.063123},
  url = {https://link.aps.org/doi/10.1103/PhysRevA.110.063123}
}

@article{Itano1982,
  author  = {Itano, W. M. and Lewis, L. L. and Wineland, D. J.},
  title   = {Shift of $^2${S}$_{1/2}$ hyperfine splittings due to blackbody radiation},
  journal = {Phys. Rev. A},
  volume  = {25},
  pages   = {1233--1235},
  year    = {1982},
  doi     = {10.1103/PhysRevA.25.1233}
}

@article{Farley1981,
  author  = {Farley, J. W. and Wing, W. H.},
  title   = {Accurate calculation of dynamic {S}tark shifts and depopulation rates
              of {R}ydberg energy levels induced by blackbody radiation:
              Hydrogen, helium, and alkali-metal atoms},
  journal = {Phys. Rev. A},
  volume  = {23},
  pages   = {2397--2424},
  year    = {1981},
  doi     = {10.1103/PhysRevA.23.2397}
}

@article{Safronova2012,
  author  = {Safronova, M. S. and Kozlov, M. G. and Clark, C. W.},
  title   = {Blackbody radiation shifts in optical atomic clocks},
  journal = {IEEE Trans. Ultrason. Ferroelectr. Freq. Control},
  volume  = {59},
  pages   = {439--447},
  year    = {2012},
  doi     = {10.1109/TUFFC.2012.2213}
}

@article{Ruggenthaler2018,
  author  = {Ruggenthaler, M. and Tancogne-Dejean, N. and Flick, J. and Appel, H. and Rubio, A.},
  title   = {From a quantum-electrodynamical light--matter description to novel spectroscopies},
  journal = {Nat. Rev. Chem.},
  volume  = {2},
  pages   = {0118},
  year    = {2018},
  doi     = {10.1038/s41570-018-0118}
}

@article{Ruggenthaler2014pra,
  author  = {Ruggenthaler, M. and Flick, J. and Pellegrini, C. and Appel, H.
             and Tokatly, I. V. and Rubio, A.},
  title   = {Quantum-electrodynamical density-functional theory: {B}ridging
             quantum optics and electronic-structure theory},
  journal = {Phys. Rev. A},
  volume  = {90},
  pages   = {012508},
  year    = {2014},
  doi     = {10.1103/PhysRevA.90.012508}
}

@article{Flick2017pnas,
  author  = {Flick, J. and Ruggenthaler, M. and Appel, H. and Rubio, A.},
  title   = {Atoms and molecules in cavities, from weak to strong coupling in
             quantum-electrodynamics ({QED}) chemistry},
  journal = {Proc. Natl. Acad. Sci. USA},
  volume  = {114},
  pages   = {3026--3034},
  year    = {2017},
  doi     = {10.1073/pnas.1615509114}
}

@article{Mandal2023,
  author  = {Mandal, A. and Taylor, M. A. and Weight, B. M. and others},
  title   = {Theoretical Advances in Polariton Chemistry and Molecular Cavity
             Quantum Electrodynamics},
  journal = {Chem. Rev.},
  volume  = {123},
  pages   = {9786--9879},
  year    = {2023},
  doi     = {10.1021/acs.chemrev.2c00855}
}

@article{Haugland2020jctc,
  author  = {Haugland, T. S. and Ronca, E. and Kj{\o}nstad, E. F. and
             Rubio, A. and Koch, H.},
  title   = {Coupled Cluster Theory for Molecular Polaritons: {C}hanging
             Ground and Excited States},
  journal = {Phys. Rev. X},
  volume  = {10},
  pages   = {041043},
  year    = {2020},
  doi     = {10.1103/PhysRevX.10.041043}
}

@article{Haugland2021jcp,
  author  = {Haugland, T. S. and Sch\"{a}fer, C. and Ronca, E. and Rubio, A. and Koch, H.},
  title   = {Intermolecular interactions in optical cavities: {A}n \textit{ab initio} {QED} study},
  journal = {J. Chem. Phys.},
  volume  = {154},
  pages   = {094113},
  year    = {2021},
  doi     = {10.1063/5.0039256}
}

@article{Riso2022,
  author  = {Riso, R. R. and Haugland, T. S. and Ronca, E. and Koch, H.},
  title   = {Molecular orbital theory in cavity {QED} environments},
  journal = {Nat. Commun.},
  volume  = {13},
  pages   = {1368},
  year    = {2022},
  doi     = {10.1038/s41467-022-29003-2}
}

@article{DePrince2021,
  author  = {DePrince, A. E., III},
  title   = {Cavity-modulated ionization potentials and electron affinities
             from quantum electrodynamics coupled-cluster theory},
  journal = {J. Chem. Phys.},
  volume  = {154},
  pages   = {094112},
  year    = {2021},
  doi     = {10.1063/5.0038748}
}

@article{Horak2025,
  author  = {Horak, J. and others},
  title   = {Analytic model reveals local molecular polarizability changes
             induced by collective strong coupling in optical cavities},
  journal = {Phys. Rev. Research},
  volume  = {7},
  pages   = {013242},
  year    = {2025},
  doi     = {10.1103/PhysRevResearch.7.013242}
}

@article{Philbin2023,
  author  = {Philbin, J. P. and others},
  title   = {Molecular van der {W}aals Fluids in Cavity Quantum Electrodynamics},
  journal = {J. Phys. Chem. Lett.},
  volume  = {14},
  pages   = {8988--8993},
  year    = {2023},
  doi     = {10.1021/acs.jpclett.3c01790}
}

@article{DePrince2025,
  author  = {DePrince, A. E., III and Yuwono, S. H.},
  title   = {Static Electric Dipole Polarizability and Hyperpolarizability
             Tensors from Mean-Field Cavity Quantum Electrodynamics
             Approaches},
  journal = {J. Phys. Chem. A},
  volume  = {129},
  pages   = {8024--8033},
  year    = {2025},
  doi     = {10.1021/acs.jpca.5c03749}
}

@article{Zalialiutdinov2025dboc,
    author = {Zalialiutdinov, T. and Solovyev, D. and Lopez-Rodriguez, J. J. and Anikin, A. and Kotov, A.},
    title = {Non-adiabatic effects induced by strong light–matter coupling in cavity QED},
    journal = {The Journal of Chemical Physics},
    volume = {165},
    number = {12},
    pages = {124104},
    year = {2026},
    month = {09},
    issn = {0021-9606},
    doi = {10.1063/5.0349306},
    url = {https://doi.org/10.1063/5.0349306}
}

@article{Schiller2014,
  author  = {Schiller, S. and Bakalov, D. and Korobov, V. I.},
  title   = {Simplest Molecules as Candidates for Precise Optical Clocks},
  journal = {Phys. Rev. Lett.},
  volume  = {113},
  pages   = {023004},
  year    = {2014},
  doi     = {10.1103/PhysRevLett.113.023004}
}

@article{Hutson2024,
  author  = {Hutson, R. B. and Milner, W. R. and Yan, L. and Ye, J. and
             Sanner, C.},
  title   = {Observation of millihertz-level cooperative {L}amb shifts in
             an optical atomic clock},
  journal = {Science},
  volume  = {383},
  pages   = {384--387},
  year    = {2024},
  doi     = {10.1126/science.adh4477}
}

@article{CP2026,
  author  = {Ton, K. and Kestler, G. and Steck, D. A. and Barreiro, J. T.},
  title   = {Spectroscopic measurement of the {C}asimir-{P}older force in
             the intermediate regime},
  journal = {arXiv preprint},
  note    = {arXiv:2604.14721},
  year    = {2026}
}

@article{Chikkaraddy2016,
  author  = {Chikkaraddy, R. and de Nijs, B. and Benz, F. and others},
  title   = {Single-molecule strong coupling at room temperature in
             plasmonic nanocavities},
  journal = {Nature},
  volume  = {535},
  pages   = {127--130},
  year    = {2016},
  doi     = {10.1038/nature17974}
}

@article{BishopCheung1978,
  author  = {Bishop, D. M. and Cheung, L. M.},
  title   = {Moment functions (including static dipole polarisabilities)
              and radiative corrections for H$_2^+$},
  journal = {J. Phys. B: At. Mol. Phys.},
  volume  = {11},
  pages   = {3133--3144},
  year    = {1978},
  doi     = {10.1088/0022-3700/11/18/010}
}

@article{Cohen1965,
  author  = {Cohen, H. D. and Roothaan, C. C. J.},
  title   = {Electric Dipole Polarizability of Atoms by the {Hartree-Fock}
             Method. {I}. Theory for Closed-Shell Systems},
  journal = {J. Chem. Phys.},
  volume  = {43},
  pages   = {S34--S39},
  year    = {1965},
  doi     = {10.1063/1.1701512}
}

@article{Sadlej1992,
  author  = {Sadlej, A. J.},
  title   = {Medium-size polarized basis sets for high-level correlated
             calculations of molecular electric properties},
  journal = {Collect. Czech. Chem. Commun.},
  volume  = {53},
  pages   = {1995},
  year    = {1988},
  doi     = {10.1135/cccc19881995}
}

@article{Baranowska2009,
  author  = {Baranowska, A. and Sadlej, A. J.},
  title   = {Polarized basis sets for accurate calculations of static and
             dynamic electric properties of molecules},
  journal = {J. Comput. Chem.},
  volume  = {31},
  pages   = {552},
  year    = {2010},
  doi     = {10.1002/jcc.21348}
}

@misc{OpenMS,
  author = {Zhang, Y. and others},
  title  = {{OpenMS}: A multiscale ecosystem for solving coupled
            {Maxwell--Schr\"{o}dinger} dynamics},
  note   = {\url{https://github.com/lanl/OpenMS}},
  year   = {2023}
}

@article{Sun2018,
  author  = {Sun, Q. and Berkelbach, T. C. and Blunt, N. S. and others},
  title   = {{PySCF}: The {Python}-based simulations of chemistry framework},
  journal = {WIREs Comput. Mol. Sci.},
  volume  = {8},
  pages   = {e1340},
  year    = {2018},
  doi     = {10.1002/wcms.1340}
}

@article{Rooij2011,
  author  = {van Rooij, R. and Borbely, J. S. and Simonet, J. and Hoogerland, M. D. and Eikema, K. S. E. and Rozendaal, R. A. and Vassen, W.},
  title   = {Frequency metrology in quantum degenerate helium: Direct measurement of the $2^3${S}$_1$ $\to$ $2^1${S}$_0$ transition},
  journal = {Science},
  volume  = {333},
  pages   = {196--198},
  year    = {2011},
  doi     = {10.1126/science.1205163}
}

@article{Notermans2014,
  author  = {Notermans, R. P. M. J. W. and Vassen, W.},
  title   = {High-precision spectroscopy of the forbidden $2^3${S}$_1$ $\to$ $2^1${P}$_1$ transition in quantum degenerate metastable helium},
  journal = {Phys. Rev. Lett.},
  volume  = {112},
  pages   = {253002},
  year    = {2014},
  doi     = {10.1103/PhysRevLett.112.253002}
}

@article{Yan2000,
  author  = {Yan, Zong-Chao},
  title   = {Polarizabilities of the {R}ydberg states of helium},
  journal = {Phys. Rev. A},
  volume  = {62},
  pages   = {052502},
  year    = {2000},
  doi     = {10.1103/PhysRevA.62.052502}
}

@article{YanBabb1998,
  author  = {Yan, Z.-C. and Babb, J. F.},
  title   = {Long-range interactions of metastable helium atoms},
  journal = {Phys. Rev. A},
  volume  = {58},
  pages   = {1247},
  year    = {1998},
  doi     = {10.1103/PhysRevA.58.1247}
}

@article{Knowles1984,
  author  = {Knowles, P. J. and Handy, N. C.},
  title   = {A new determinant-based full configuration interaction method},
  journal = {Chem. Phys. Lett.},
  volume  = {111},
  pages   = {315--321},
  year    = {1984},
  doi     = {10.1016/0009-2614(84)85513-X}
}

@article{Perspective2025,
  author  = {Bauman, N. and others},
  title   = {Perspective on Many-Body Methods for Molecular Polaritonic
             Systems},
  journal = {J. Chem. Theory Comput.},
  volume  = {21},
  pages   = {10035--10067},
  year    = {2025},
  doi     = {10.1021/acs.jctc.5c00801}
}

@article{eT2020,
  author  = {Folkestad, Sarai D. and Kjønstad, Eirik F. and Myhre,
            Rolf H. and Andersen, Josefine H. and Balbi, Alice and
            Coriani, Sonia and Giovannini, Tommaso and Goletto, Linda
            and Haugland, Tor S. and Hutcheson, Anders and Høyvik,
            Ida-Marie and Moitra, Torsha and Paul, Alexander C. and
            Scavino, Marco and Skeidsvoll, Andreas S. and Tveten,
            Asmund H. and Koch, Henrik},
  title   = {{eT} 1.0: An open source electronic structure program
             with emphasis on coupled cluster and multilevel methods},
  journal = {J. Chem. Phys.},
  volume  = {152},
  pages   = {184103},
  year    = {2020},
  doi     = {10.1063/5.0004713}
}

\end{document}